\documentclass[12pt]{article}
\pdfoutput=1
\usepackage{amsbsy}
\usepackage{amsfonts}
\usepackage{amsmath,amssymb}
\usepackage{bbold}
\usepackage{pdfpages}

\newcommand{\sect}[1]{\setcounter{equation}{0}\section{#1}}

\newcommand{\eq}{\begin{equation}}
\newcommand{\eqa}{\begin{eqnarray}}  
\newcommand{\en}{\end{equation}}
\newcommand{\ena}{\end{eqnarray}}
\newcommand{\enn}{\nonumber \end{equation}}

\def\sk{\vskip .4cm}
\def\noi{\noindent}
\def\om{\omega}

\def\ga{\gamma}

\let \part\partial

\def\unmezzo{{1 \over 2}}
\def\epsi{\varepsilon}
\def\we{\wedge}

\def\de{\delta}

\def\part{\partial}

\def\sk{\vskip .4cm}

\def\noi{\noindent}

\def\X0{X^0}

\def\om{\omega}

\def\ga{\gamma}

\def\unmezzo{{1 \over 2}}
\def\epsi{\varepsilon}

\def\we{\wedge}

\def\de{\delta}

\def\Dcal{{\cal D}}

\def\Mcal{{\cal M}}

\def\square{{\,\lower0.9pt\vbox{\hrule \hbox{\vrule height 0.2 cm
\hskip 0.2 cm \vrule height 0.2 cm}\hrule}\,}}

\def\westar{\we_\star}

\def\omtilde{\widetilde \om}
\def\Vtilde{\widetilde{V}}

\def\rtilde{\tilde r}
\def\epsitilde{\widetilde{\epsi}}

\def\Om{\Omega}

\def\Gbb{\mathbb{G}}

\def\dright{\stackrel{\rightarrow}{\part}}
\def\dleft{\stackrel{\leftarrow}{\part}}

\def\pitilde{{\widetilde \pi}}
\def\Phitilde{{\widetilde \Phi}}

\begin{document}

\begin{titlepage}
\vskip 2em
\begin{center}
{\Large \bf Noncommutative hamiltonian formalism \\ \sk for noncommutative gravity}
\vskip 0.5cm

{\bf
Leonardo Castellani}
\medskip

\vskip 0.5cm

{\sl Dipartimento di Scienze e Innovazione Tecnologica
\\Universit\`a del Piemonte Orientale, viale T. Michel 11, 15121 Alessandria, Italy\\ [.5em] INFN, Sezione di 
Torino, via P. Giuria 1, 10125 Torino, Italy\\ [.5em]
Regge Center for Algebra, Geometry and Theoretical Physics, via P. Giuria 1, 10125 Torino, Italy
}\\ [4em]
\end{center}

\begin{abstract}
\sk

We present a covariant canonical formalism for noncommutative (NC) gravity, and in general for
NC geometric theories defined via a twisted $\star$-wedge product between forms. 
Noether theorems are generalized to the NC setting, and gauge generators 
are constructed in a twisted phase space with $\star$-deformed Poisson bracket. This formalism is applied to NC $d=4$ vierbein gravity, and
allows to find the canonical  generators of the tangent space $\star$-gauge group. 

\end{abstract}
\vskip 3cm

{\it ~~~~~~~~~~~~~~~~~~~~~~~In memoriam of Alessandro D' Adda}

\vskip 5cm
 \noi \hrule \vskip .2cm \noi {\small
leonardo.castellani@uniupo.it}

\end{titlepage}

\newpage
\setcounter{page}{1}

\tableofcontents

\vfill\eject

\sect{Introduction}

In this paper we present a hamiltonian formalism tailored for noncommutative (NC) geometric field theories. Both hamiltonian techniques and noncommutative geometry have a long history in theoretical physics, and have been brought together already many decades ago when phase space, the geometrical arena of canonical formalism, was studied by Dirac \cite{Dirac1925,Dirac1926} as a primordial example of noncommutative space. This inspired the idea of {\sl spacetime coordinates} as non-commuting operators \cite{Snyder1,Snyder2}, with commutation relations of the type $[x^\mu,x^\nu] = i \theta^{\mu\nu}$. Such relations lead to an uncertainty principle that smears spacetime at distances shorter than $\sqrt{\theta}$. Since quantum theory prevents us to ``measure" geometry at distances smaller than the Planck length $L_P$ (at this scale the curvature radius of spacetime becomes comparable to the wavelength of a probe particle), the 
above commutation relations seem to make good physical sense when $\sqrt{\theta} \approx L_P$. A quantum theory of gravity containing or predicting
noncommutativity of spacetime coordinates at small distances would therefore stand a good chance to be intrinsically regulated.

Geometric theories, as for example (super)gravity, can be formulated on a NC spacetime,
 by deforming the usual wedge product between forms to a twisted wedge product, or $\wedge_\star$ product. This program has been applied for instance to Yang-Mills theories coupled to fermions (for a NC version of the standard model see for ex. \cite{Calmet2001,Aschieri2002}), to gravity coupled to fermions \cite{AC1}, and also to supergravity, see \cite{AC2}.

Canonical quantization of these NC field theories can be achieved by applying the traditional hamiltonian methods to the classical
$\star$-deformed theories, expanded in the noncommutativity parameter $\theta$. Another way, which we advocate in the present paper, is to design a NC hamiltonian formalism that can be applied directly to the NC theory, before expressing it as a classical deformed field theory.

  Here we generalize to a NC setting the {\sl covariant} hamiltonian formalism, introduced in the eighties in ref.s \cite{CCF1}-\cite{CCF5}, and further developed more recently in \cite{CD,C2020}. This formalism is well adapted to geometric theories, governed by Lagrangian $d$-forms.
    
 By extending the covariant Legendre transformation of \cite{CD,C2020} to a NC covariant Legendre transformation, we can define NC momenta and Hamiltonian. This we do explicitly on the example of the twisted noncommutative (pure) gravity in $d=4$ of ref. \cite{AC1}.
 
 The power of hamiltonian methods is particularly useful when investigating symmetries. Starting from the Hamiltonian, an algorithm \cite{SCHS} exists to construct all the canonical symmetry generators, based on the Dirac treatment of constrained hamiltonian systems \cite{Dirac,HRT,HT}. The algorithm has been generalized to the covariant hamiltonian framework in \cite{CD,C2020}, and in the present paper further generalized to NC theories, thus providing a NC analogue of Noether theorems.
 
 The paper is organized as follows. Section 2 summarizes the covariant hamiltonian method for geometric theories. Section 3 generalizes this method to NC twisted geometric theories. In Sections 4 and 5 
 we give a detailed account of $d=4$ NC gravity in this formalism: NC momenta, Hamiltonian and Poisson brackets are found, and the canonical generator of the NC symmetries of the theory is explicitly constructed via the NC generalization of the algorithm of \cite{CD,C2020}. The Appendices contain a micro-review of twisted NC geometry (for a more detailed treatment see for ex. \cite{book}), and some gamma matrix conventions.

 \sect{A summary of the covariant hamiltonian formalism} 

 \subsection{Geometric action and field equations}
 
 Consider the action :
\eq
S= \int_{\Mcal^d} L (\phi_i, d\phi_i) \label{geomaction}
\en
where the Lagrangian $L$ depends on a collection of $p_i$-form fields $\phi_i$ and their exterior derivatives, and is
integrated on a $d$-dimensional manifold $\Mcal^d$.

The variational principle reads
\eq
\delta S =  \int_{\Mcal^d} \delta \phi_i { \dright L \over \partial \phi_i} + d (\delta \phi_i ) { \dright L \over \partial (d\phi_i)}=0  \label{variational}
\en
All products are exterior products between forms, satisfying
\eq
A B = (-)^{ab + \eta_a \eta_b} BA
\en
with $a,b$ and $\eta_a,\eta_b$ the degrees and the fermionic gradings of the forms $A$ and $B$ respectively
($\eta = 0$ for bosons and $\eta = 1$  for fermions). In the following we consider for simplicity only bosonic fields. The symbol ${ \dright L \over \partial \phi_i}$ indicates the right derivative of $L$ 
 with respect to a $p_i$-form $\phi_i$, defined by first bringing $\phi_i$ to the left in $L$
(taking into account the sign changes due to the form degrees)
and then canceling it against the derivative. 

After integrating by parts the second term in the action variation (\ref{variational}), one obtains the Euler-Lagrange equations
\eq
d ~ { \dright L \over \partial (d\phi_i)} - (-)^{p_i} { \dright L \over \partial \phi_i} =0 \label{ELeqs}
\en
These are $(d-p_i)$-form equations.

\subsection{Form Hamiltonian}

We define the $d$-form Hamiltonian as:
\eq
H \equiv d\phi_i ~\pi^i  - L \label{formH}
\en
where the ($d-p_i-1$)-form momenta are given by:
\eq
\pi^i \equiv {\dright L \over \partial (d\phi_i)} \label{momentadef}
\en
The Hamiltonian does not depend on the ``velocities" $d\phi_i$, the proof being the same as in the usual
case, unless primary constraints are present (see later).

The form-analogue of the Hamilton equations reads:
\eq
d \phi_i = (-)^{(d+1)(p_i+1)} ~ {\dright H \over \partial \pi^i} ,~~~d \pi^i =  (-)^{p_i+1} ~{\dright H \over \partial \phi_i} \label{formHE}
\en
The first equation is equivalent to the momentum definition, and is obtained by taking the right derivative of $H$ as given in 
(\ref{formH}) with respect to $\pi^i$, and then using (\ref{momentadef}), and $(d-p_i-1)(p_i+1) = (d+1)(p_i + 1) (mod~ 2)$. The second is equivalent to the Euler-Lagrange form equations, and is obtained by taking the right derivative of $H$ with respect to $\phi_i$. Both can be deduced as variational equations for the action
(\ref{geomaction}) rewritten with canonical variables:
\eq
S =  \int_{\Mcal^d} d\phi_i ~\pi^i  - H(\phi_i, \pi^i)
\en

\subsection{Form Poisson bracket}

The form Hamilton equations allow to express the (on shell) exterior differential of any $p$-form $F(\phi_i, \pi^i)$ as
\eq
dF=d\phi_i ~{\dright F\over \partial \phi_i} + d \pi^i ~{\dright F\over \partial \pi^i} =  (-)^{(d+1)(p_i+1)} ~{\dright H \over \partial \pi^i}~
{\dright F\over \partial \phi_i} +  (-)^{p_i+1} ~{\dright H \over \partial \phi_i} {\dright F\over \partial \pi^i} \label{dF}
\en
Using left derivatives this expression simplifies:
\eq
dF= {\dleft H \over \partial \pi^i}~
{\dright F\over \partial \phi_i} -  (-)^{p_i d} ~{\dleft H \over \partial \phi_i} {\dright F\over \partial \pi^i}  \label{differential}
\en
{\bf Note:} left derivatives are defined as ``acting on the left" and for example ${\dleft H \over \partial \phi_i}$ really means
$H {\dleft \over \partial \phi_i}$. It is easy to verify that the left and right derivatives of an $f$-form $F$ with respect 
to an $a$-form $A$ satisfy
\eq
{\dleft F \over \partial A} = (-)^{a(f+1)} ~{\dright F \over \partial A} \label{leftright}
\en
and this relation is used to prove eq. (\ref{differential}).
\sk

The expression for the differential (\ref{differential}) suggests the definition of the {\it form Poisson bracket} (FPB):
\eq
\{ A, B \} \equiv  {\dleft B \over \partial \pi^i}~
{\dright A\over \partial \phi_i} -  (-)^{p_i d} ~{\dleft B \over \partial \phi_i} {\dright A\over \partial \pi^i}  \label{FPB}
\en
so that
\eq
dF = \{ F,H \} \label{differential2}
\en
The form Poisson bracket between the $a$-form $A$ and the $b$-form $B$ is a ($a+b-d+1$)-form, and canonically conjugated forms satisy:
\eq
 \{ \phi_i, \pi^j\}  = \delta_i^j \label{canonicalPB}
 \en

Using the definition (\ref{FPB}), the following relations for the FPB of  (\ref{FPB}) can be shown to hold \cite{CD}:
\eqa
& & \{ B,A \} = - (-)^{(a+d+1)(b+d+1)} \{ A,B \}  \label{prop1} \\
& & \{A,BC \} = B \{A,C \} + (-)^{c(a+d+1)} \{A,B \} C  \label{prop2}\\
& & \{AB,C \} =  \{A,C \} B + (-)^{a(c+d+1)}  A \{B,C \}  \label{prop3} \\
& & (-)^{(a+d+1)(c+d+1)} \{ A, \{ B,C \} \} + cyclic~=0\\
& & (-)^{(a+d+1)(b+d+1)} \{  \{ B,C \},A \} + cyclic~=0 \label{prop5}
\ena
i.e. graded antisymmetry, derivation property, and form-Jacobi identities.  

As discussed in \cite{CD,C2020}, canonical transformations, symmetry generators and 
Noether theorems have their covariant counterpart. We will treat them directly in the 
twisted case.

  \sect{Twisted hamiltonian formalism}

In this Section we extend the covariant hamiltonian formalism to geometric theories
on noncommutative spaces, and more precisely to twisted geometric theories 
with a noncommutative $\star$-product. Many formulae of the preceding Section continue to hold, with the usual exterior product replaced by the $\star$-exterior product between forms.

\subsection{$\star$-Hamiltonian}

Starting from a geometric action
\eq
S= \int_{\Mcal^d} L (\phi_i, d\phi_i, \wedge_\star) \label{geomactionstar}
\en
where all the fields are multiplied with $\wedge_\star$ products, the Euler-Lagrange equations
are obtained by varying with respect to $\phi_i$. This variation takes the same form as in the
classical (undeformed) case (\ref{variational}), with two modifications:
\sk

\noi $\bullet$  exterior products $\wedge$ are replaced by $\wedge_\star$.

\noi $\bullet$  the right derivative is defined via cyclic permutation. This means that the field to be varied is first
brought to the left via cyclic permutation (this we can do since integration is cyclic), and then is varied.
\sk
The outcome is that the Euler-Lagrange equations take the same form as the classical ones, with
the right derivative being defined with the cyclic procedure. 
\sk
The $\star$-Hamiltonian $d$-form is defined as
\eq
H \equiv d\phi_i  \wedge_\star \pi^i  - L \label{formstarH}
\en
where the ($d-p_i-1$)-form $\star$-momenta are given by:
\eq
\pi^i \equiv {\dright L \over \partial (d\phi_i)} \label{starmomentadef}
\en
with cyclic right derivative. Note that here only the {\sl integral} of $H$ is independent of $d\phi_i$, since
cyclicity is essential for the proof, and only integrated $\wedge_\star$ products are cyclic. 

Varying the action
\eq
S =  \int_{\Mcal^d} d\phi_i ~ \wedge_\star \pi^i  - H  \label{staractionH}
\en
yields the $\star$--Hamilton equations: these are formally the same as the classical ones in (\ref{formHE}),
with the proviso that the right derivative is the cyclic one. 

In the following right and left derivatives are always defined
to be cyclic, and $\wedge_\star$ products between forms are understood when omitted.

\subsection{$\star$-Poisson bracket and infinitesimal canonical transformations}

Formula (\ref{dF}), with cyclic derivatives and $\wedge_\star$ products, holds now only under integration, so that $F$ must be a ($d-1$)-form. Still, the
relation (\ref{leftright}) continues to hold, with cyclic left and right derivatives, and we can define
$\star$-Poisson brackets formally as in (\ref{FPB}):
\eq
\{ A, B \}_\star \equiv  {\dleft B \over \partial \pi^i} \wedge_\star
{\dright A\over \partial \phi_i} -  (-)^{p_i d} ~{\dleft B \over \partial \phi_i} \wedge_\star {\dright A\over \partial \pi^i}  \label{starFPB}
\en

As in the classical case, we can also define infinitesimal canonical transformations
 on the basic fields and momenta as follows:
 \eqa
  & & \delta \phi_i = \{ \phi_i, G \}_\star = {\dleft G\over \partial \pi^i} \nonumber \\
    & & \delta \pi^i = \{ \pi^i,G \}_\star = - (-)^{p_i d}  {\dleft G\over \partial \phi_i}  \label{ICT}
 \ena
 where $G=G(\phi_i, \pi^i)$ is a ($d -1$)--form, the generator of the canonical transformation. 
The generator $G$ can contain parameters $\epsi (x)$ depending only on spacetime, for example as $G = \epsi (x)  G'$. 
 For a $d$-form $A$, the following relation holds under integration:
 \eq
 \delta A = \delta \phi_i ~ {\dright A \over \partial \phi_i} +  \delta \pi^i ~{\dright A \over \partial \pi^i} = \{A,G\}_\star  \label{deltaA}
 \en
The second equality follows immediatedly from eq.s (\ref{ICT}). 
 Note that $ \{A,G \}_\star$ is a $d$-form like $A$. 
 
 In the following we always consider ($d$-1)--form generators
 of the type $\epsi (x)~ \pi \phi$ (``point canonical transformations“ )  or of the
 more general form $d\epsi (x) ~\pi + \epsi (x)~ \pi \phi $, and it is easy to check that 
the transformations (\ref{ICT}) preserve the canonical FPB relations (\ref{canonicalPB}). Indeed
\eqa
& & \{\phi'_i, \pi'^j \}_\star = \{\phi_i +  \{\phi_i ,G\}_\star, \pi^j +  \{ \pi^j,G \}_\star \}_\star  \nonumber \\
& & ~~~~~~~~~~~ = \{\phi_i , \pi^j  \}_\star  +   \{\phi_i ,  \{ \pi^j ,G \}_\star \}_\star +   \{ \{ \phi_i,G  \}_\star , \pi^j \}_\star  + O(\epsi^2) \nonumber \\
& & ~~~~~~~~~~~ = \{\phi_i , \pi^j  \}_\star  + O(\epsi^2)
\ena
since the two central terms of the second line cancel each other, as can be verified using the definition of $\star$-Poisson bracket (\ref{starFPB}). Therefore the variations (\ref{ICT}) deserve to be called $\star$-{\sl canonical (infinitesimal) transformations}. 

\subsection{$\star$-Symmetries and $\star$-Noether theorems}

Consider now the generic variation of the action (\ref{staractionH}) 
\eqa
& & \delta S = \int_{\Mcal^d} d (\delta \phi_i ) \pi^i + d \phi_i ~\delta \pi^i - \delta H
 = \int_{\Mcal^d} d (\delta \phi_i  \pi^i )- (-)^{p_i} \delta \phi_i ~d\pi^i+ d \phi_i ~\delta \pi^i - \delta H \nonumber \\
& & ~~~~ =\int_{\Mcal^d} d (\delta \phi_i  \pi^i )- (-)^{p_i} (-)^{(d-p_i)p_i} d\pi^i ~\delta \phi_i + d \phi_i ~\delta \pi^i - \delta H 
\ena
where  $\wedge_\star$
products are omitted. Recalling the variations (\ref{ICT}), and the relation (\ref{leftright}) between right and left derivatives, the variation of the action becomes
\eqa
& & \delta S = \int_{\Mcal^d}  d({\dleft G\over \partial \pi^i} \pi^i ) - d \pi^i {\dright G\over \partial \pi^i} - d\phi_i {\dright G\over \partial \phi_i}  - \{H,G \}_\star= \nonumber \\
& & ~~~~=\int_{\Mcal^d} d({\dleft G\over \partial \pi^i} \pi^i  -G) - \{H,G \}_\star
\ena
 where we have used (\ref{deltaA}) for the variation of $H$, since $H$ is a $d$--form. 
 In the final expression for $\delta S$ we have equated $dG$ to the two terms with right derivatives:
 this is correct as long as $G$ contains only $\phi$ and $\pi$ fields, plus possibly constant (infinitesimal)
 parameters $\epsi$. If it contains arbitrary functions $\epsi (x)$, for example $\Gbb=\epsi (x) G$ (we denote
 in this case the generator with $\Gbb$, while $G$ is still a function of only $\phi$ and $\pi$), the two terms with right derivatives become $-d(\epsi (x) G) + d \epsi(x) ~G$ under integration. With appropriate conditions on the
 boundary of ${\Mcal^d}$, the total derivative can be ignored and we are left with:
 \eq
 \delta S = \int_{\Mcal^d} d \epsi~ G - \{H, \epsi G \}_\star  \label{variationS}
 \en
 For constant $\epsi$, we recover the analogue of Noether's theorem for global symmetries: the symmetry generator $\Gbb=\epsi G$ has vanishing $\star$--Poisson bracket with the Hamiltonian. For constant $\epsi$
 this also implies $\{H,G\}_\star =0$.  As in the classical case, this leads to a conservation law, since the on-shell differential of $G$ can be expressed (under integration) as $dG= \{ G,H \}_\star = - \{H,G \}_\star$. 
 Then on shell $\int_{\Mcal^d} dG =0$ and the integral of $G$ on the boundary ${\partial \Mcal^d}$ vanishes.
 Taking vanishing fields at spatial infinity, the integral of $G$ on a spacelike surface is conserved in time.
 
 When $\epsi$ is nonconstant, the variation (\ref{variationS}) vanishes iff $G=0$ and $\{H, \epsi G\}_\star=0$
 (a nonconstant $\epsi$ cannot be taken outside the $\star$--Poisson bracket, since it $\star$--multiplies $G$).
 Thus generators of local symmetries have to vanish, together with their $\star$--Poisson bracket with the Hamiltonian, a result analogous to Noether's theorem for local symmetries.
 
 In gauge and gravity theories the symmetry transformations on the basic fields can include 
 derivatives of the parameter. In our geometric formalism these can only be exterior derivatives $d\epsi$.
 Therefore we must consider generators $\Gbb$ of the type
 \eq
  \Gbb = \epsi G + d \epsi ~F \label{generatorG}
  \en
 where $F$ is a ($d-2$)--form, and the action variation becomes
 \eq
  \delta S = \int_{\Mcal^d} d \epsi~ G - \{H,\epsi G \}_\star - \{H, d \epsi F \}_\star \label{variationS2}
 \en
 Thus $\star$-symmetry generators of this type must satisfy separately
 \eq
 \{H,\epsi G \}_\star = 0,~~~ d \epsi~ G - \{H,d \epsi F \}_\star = 0  \label{conditions}
 \en
since $\epsi$ is an arbitrary function. To check whether these conditions (\ref{conditions}) hold
we can use cyclic reorderings, since they originate from the varied integral (\ref{variationS2}).
Note that in presence of hamiltonian constraints $\Phi =0$ (as is the case when the action 
is invariant under gauge-type transformations) one must also require the variation of the constraints to vanish weakly, i.e. modulo constraints:
\eq
\delta \Phi \approx 0   \label{conditions2}
\en
with $\approx$ meaning weak equality.  In the commutative case, these variations can be expressed with the Poisson bracket $\{ \Phi,\Gbb \}$, whereas in the NC setting $\delta \Phi = \{ \Phi,\Gbb \}_\star$ holds only under integration, with $\Phi$  a $d$--form. 
\sk
  \noi This generalizes to the noncommutative setting the algorithm of
\cite{SCHS} (adapted to the covariant hamiltonian formalism in \cite{CD}) for the construction of canonical gauge symmetry generators. The steps of the algorithm are:
\sk
1) look for first class $(d-1)$--forms $d\epsi F (\phi, \pi)$, first class meaning that they have weakly vanishing $\star$-Poisson bracket with all the constraints.

 2) compute the $\star$--Poisson bracket $\{H,d \epsi F \}_\star$. The result leads to a candidate for $G$.

 3) Check whether this candidate, possibly with the addition of weakly vanishing pieces, satisfies
$ \{H,\epsi G  \}_\star = 0$. In fact this request usually fixes the weakly vanishing pieces (combinations of constraints) that need to be added. 

4) Check whether the variations generated by $\Gbb=\epsi G + d \epsi ~F$ preserve (weakly) the
constraints $\Phi$.

Then $\Gbb$ generates a symmetry of the action (\ref{staractionH}). 

\sk
In the next Sections we apply this formalism to noncommutative (twisted) $d=4$ vierbein gravity.

  \sect{Twisted noncommutative gravity in $d=4$}
 
 \subsection{Classical action and symmetries}
 
 We start by rewriting the classical Einstein-Hilbert action in a compact form, using an index-free notation:

\eq 
S =  \int Tr \left(i R \we V \we V \ga_5 \right) \label{action1}
\en

\noi The fundamental fields are the 1--forms $\Om$  (spin
connection) and $V$ (vierbein):
\eq 
 \Om = {1 \over 4} \om^{ab} \ga_{ab}, ~~~~~V = V^a
\ga_a ~~~~~~\label{VaOmab}
\en
The curvature 2--form $R$ is defined as
 \eq
  R= d\Om - \Om \we
\Om, 
\en
\noi Thus all fields are $4 \times 4$ matrices in the spinor representation
 (see Appendix A for $d=4$ gamma matrix conventions).
 The trace $Tr$ is taken on this representation.
Using the $d=4$ gamma matrix  identities
\eq
\ga_{abc} = i \epsi_{abcd} \ga^d \ga_5,~~~~~~~
 Tr (\ga_{ab} \ga_c \ga_d \ga_5) = -4 i \epsi_{abcd}
  \en
 \noi we recover the usual action
 \eq
 S = \int R^{ab} \we V^c \we V^d ~\epsi_{abcd}   \label{EHaction}
 \en

The action is invariant under local diffeomorphisms
 (it is the integral of a 4-form on a 4-manifold)
  and under the local Lorentz rotations:
\eq
\de_\epsilon V = -[V,\epsilon ] , ~~~\de_\epsilon \Om = d\epsilon - [\Om,\epsilon], \label{gaugevariations}
\en  
\noi with gauge parameter (0-form)
\eq
 \epsilon = {1\over 4} \epsilon^{ab} \ga_{ab}  \label{epsiabcd}
  \en
  The invariance can be directly checked on the action (\ref{action1}) noting that
   \eq
  \de_\epsilon  R = - [{ R},\epsilon ] ,
     \en
   \noi and using the cyclicity of the trace $Tr$, and the fact that $\epsilon$ commutes with
   $\ga_5$. The Lorentz rotations close on the Lie algebra:
   \eq
   [\de_{\epsilon_1},\de_{\epsilon_2}] = -\de_{[\epsilon_1,\epsilon_2]}
   \en
   After substituting (\ref{VaOmab}) and (\ref{epsiabcd}) into (\ref{gaugevariations}), simple gamma algebra yields the gauge variations of the component fields in (\ref{VaOmab}):
   \eq
   \delta_\epsilon V^a= \epsilon^a_{~b} V^b,~~~\delta_\epsilon \omega^{ab} = d\epsilon^{ab} -\omega^a_{~c} \epsilon^{cb} + \omega^b_{~c} \epsilon^{ca}  \equiv \Dcal \epsilon^{ab}
   \en
   Similarly, the variation of the curvature components is found to be
   \eq
   \delta R^{ab} = \epsilon^{a}_{~c} R^{cb} -  \epsilon^{b}_{~c} R^{ca}
   \en
  Thus all quantities in the action (\ref{EHaction}) transform homogeneously under Lorentz
  local rotations, and since $\epsi_{abcd}$ is an invariant tensor of $SO(1,3)$, the action is
  likewise invariant. Here the proof of invariance looks simple both in the index-free and in the component 
  formulation. Note however that in general the index-free proof is much simpler.

 \subsection{NC Action and NC symmetries}

In this subsection we recall the noncommutative generalization of the action (\ref{action1}),  the bosonic part of the NC action studied in ref. \cite{AC1}. It is obtained by replacing
exterior products by deformed exterior products:

\eq S =  \int Tr \left(i {R}\westar V \westar V \ga_5  \right) \label{action1NC}
\en

\noi with
 \eq R= d\Om - \Om \westar \Om,
 \en

Almost all formulae in Section 4.1 continue to hold, with
$\star$--products and $\star$--exterior products. However, the
expansion of the fundamental fields on the Dirac basis of gamma
matrices must now include new contributions:
 \eq
  \Om = {1 \over 4} \om^{ab} \ga_{ab} + i \om 1 + \omtilde \ga_5, ~~~~~V = V^a
\ga_a + \Vtilde^a \ga_a \ga_5  ~~~~~~\label{OmVexpansions}
\en
\noi Similarly for the curvature :
 \eq
 R= {1\over 4} R^{ab} \ga_{ab} + i r 1 + \rtilde \ga_5 \label{Rexpansion}
  \en
 \noi and for the gauge parameter:
 \eq
 \epsilon = {1\over 4} \epsi^{ab} \ga_{ab} + i \epsi 1 + \epsitilde \ga_5 \label{epsiexpansion}
  \en
  \noi Indeed now the $\star$--gauge variations read:
  \eq
\de_\epsilon V = -V \star \epsilon + \epsilon \star V,
~~~\de_\epsilon \Om = d\epsilon - \Om \star \epsilon+ \epsilon
\star \Om,
\label{stargauge}\en
 \noi and in the variations for $V$ and $\Om$ also anticommutators of gamma matrices appear,
 due to the noncommutativity of the $\star$--product. Since for example the anticommutator
 $\{ \ga_{ab},\ga_{cd} \}$ contains $1$ and $\ga_5$, we see that the corresponding fields
 must be included in the expansion of $\Om$. Similarly, $V$ must contain a $\ga_a \ga_5$ term due
 to $\{ \ga_{ab},\ga_{c} \}$. Finally, the composition law for gauge parameters becomes:
 \eq
   [\de_{\epsilon_1},\de_{\epsilon_2}] = \de_{\epsilon_2 \star \epsilon_1 -
   \epsilon_1 \star \epsilon_2 }
   \en
   \noi so that $\epsilon$ must contain the $1$ and $\ga_5$ terms, since they appear in the
   composite parameter $\epsilon_2 \star \epsilon_1 - \epsilon_1 \star \epsilon_2$.

   The invariance of the noncommutative action (\ref{action1NC}) under the $\star$-variations is
   proved in exactly the same way as for the commutative case, noting that
   \eq
  \de_\epsilon R = - R \star \epsilon+ \epsilon \star R, 
     \en
   \noi and using now, besides the cyclicity of the trace $Tr$ and the fact that
   $\epsilon$ still commutes with $\ga_5$, also the graded cyclicity
   of the integral. The local $\star$-symmetry satisfies the Lie algebra of $GL(2,C)$, and centrally extends 
   the $SO(1,3)$ Lie algebra of the commutative theory.
   
   Invariance under diffeomorphisms $x \rightarrow x'(x)$ holds since the action is a 4-form integrated on 4-dimensional spacetime. More precisely, the action (\ref{action1NC}) is invariant under 
   diffeomorphisms generated by the Lie derivative ${\ell}_\epsi$, where $\epsi = \epsi^\mu \part_\mu$ is an infinitesimal tangent vector. Indeed using the Cartan formula ${\ell}_\epsi = d \iota_\epsi + \iota_\epsi d$ we find that the variation of the action is a total derivative, since the Lagrangian is a top form.
   The (infinitesimal) variation obeys the Leibniz rule, amounting to vary in turn all the fields (forms and tangent vectors) present in the action. Form fields $\tau$ vary with the Lie derivative, i.e. $\delta_\epsi \tau = \ell_\epsi \tau$, and the tangent vectors 
   $X_A= X_A^\mu \part_\mu$ defining the $\wedge_\star$ product (see Appendix A) transform as $\delta_\epsi X_A=[\epsi,X_A]$. With these variations the action varies into a total derivative\footnote{In the present setting, the tangent fields $X_A$ are background fields. Their promotion to dynamical fields is 
   discussed in ref. \cite{AC3}.}.
         
   Finally, in the next two subsections we give the NC action in terms of the component fields
     ($V^a$, $\om^{ab}$, $\Vtilde^a$, $\om$, $\omtilde$), and the gauge variations of these fields.

\subsubsection{NC gravity action for the component fields}

Using the expansions (\ref{OmVexpansions}) and (\ref{Rexpansion}), the action (\ref{action1NC}) takes the form:
\eqa
 S = & &  \int R^{ab} \westar (V^c \westar V^d - \Vtilde^c \westar \Vtilde^d)
 \epsilon_{abcd} - 2i~ R^{ab} \westar (V_a \westar  \Vtilde_b - \Vtilde_a \westar
 V_b)\nonumber \\
 & &- 4~r \westar(V^a \westar \Vtilde_a - \Vtilde^a \westar
 V_a)  + 4i~ \rtilde \westar (V^a \westar V_a -\Vtilde^a \westar
 \Vtilde_a)\nonumber \\ \label{action2NC}
\ena

\noi with
 \eqa
  & & R^{ab}= d \om^{ab} - \unmezzo \om^{a}_{~c} \westar \om^{cb} +
  \unmezzo \om^{b}_{~c} \westar \om^{ca} - i(\om^{ab}
 \westar \om + \om \westar \om^{ab}) - \nonumber \\
  & &~~~~~~  - {i \over 2}  \epsilon^{ab}_{~~cd}( \om^{cd} \westar \omtilde +
   \omtilde \westar \om^{cd}) \\
 & &   r = d\om - {i \over 8} \om^{ab} \westar \om_{ab} -i ( \om \westar
   \om - \omtilde \westar \omtilde )\nonumber \\
    & &  \rtilde = d\omtilde+ {i \over 16} \epsi_{abcd} \om^{ab} \westar
    \om^{cd}  -i (\om \westar \omtilde + \omtilde
    \westar \om) 
     \ena

\subsubsection{$\star$-gauge variations}

The action (\ref{action2NC}) is invariant under the $\star$-gauge transformations
\eqa
 & & \de_\epsi V^a = \unmezzo (\epsi^{a}_{~b} \star V^b + V^b \star
 \epsi^{a}_{~b}) + {i \over 4} \epsilon^{a}_{~bcd} (\Vtilde^{b}
 \star \epsi^{cd} -  \epsi^{cd} \star \Vtilde^{b}) \nonumber \\
& &~~~~~~~~ + i(\epsi \star V^a - V^a \star \epsi) - \epsitilde \star
\Vtilde^a - \Vtilde^a \star \epsitilde \label{NCvariations1} \\
 & & \de_\epsi \Vtilde^a = \unmezzo (\epsi^{a}_{~b} \star \Vtilde^b + \Vtilde^b \star
 \epsi^{a}_{~b}) + {i \over 4} \epsilon^{a}_{~bcd} (V^{b}
 \star \epsi^{cd} -  \epsi^{cd} \star V^{b}) \nonumber \\
& &~~~~~~~~ + i(\epsi \star \Vtilde^a - \Vtilde^a \star \epsi )-
\epsitilde \star V^a - V^a \star \epsitilde\\
 & & \de_\epsi \om^{ab} = d \epsi^{ab} +\unmezzo (\epsi^a_{~c} \star \om^{cb} -\epsi^b_{~c} \star \om^{ca}
   + \om^{cb} \star  \epsi^a_{~c} - \om^{ca} \star \epsi^b_{~c})
   \nonumber \\
   & & ~~~~~~~~ + i (\epsi^{ab} \star \om - \om \star
   \epsi^{ab}) + {i \over 2} \epsilon^{ab}_{~~cd} (\epsi^{cd} \star
   \omtilde - \omtilde \star \epsi^{cd})\\
   & & ~~~~~~~~+ i(\epsi \star \om^{ab} - \om^{ab} \star \epsi)
   + {i \over 2} \epsilon^{ab}_{~~cd} (\epsitilde \star
   \om^{cd} - \om^{cd} \star \epsitilde)\\
   & & \de_{\epsi} \om = d \epsilon - {i\over 8} (\om^{ab} \star \epsi_{ab} -
   \epsi_{ab} \star \om^{ab}) + i(\epsi \star \om - \om \star \epsi
   - \epsitilde \star \omtilde + \omtilde \star \epsitilde)\\
      & & \de_{\epsi} \omtilde = d\epsitilde +{i \over 16} \epsilon_{abcd}
    (\om^{ab} \star \epsi^{cd} - \epsi^{cd} \star \om^{ab}) +
   i( \epsi \star \omtilde - \omtilde \star \epsi + \epsitilde \star
    \om - \om \star \epsitilde)\nonumber \\ \label{NCvariations2}
    \ena
obtained by inserting the expansions (\ref{OmVexpansions}), (\ref{Rexpansion}) and (\ref{epsiexpansion}) into the variations (\ref{stargauge}).

\sect{NC Hamiltonian formulation of NC gravity}

In this Section we apply the $\star$-Hamiltonian formalism to the NC gravity action (\ref{action2NC}).

\subsection{$\star$-Legendre transformation and $\star$-Hamiltonian}

We begin by giving the definition of momenta:

\eqa
& & \pi_a =  { \dright L \over \partial (d V^a)} = 0 \label{pi1}\\
& & \pitilde_a =  { \dright L \over \partial (d \Vtilde^a)} = 0 \\
& & \pi_{ab} = {\dright L \over \partial (d \om^{ab})} =\epsilon_{abcd} (V^c V^d -\Vtilde^c \Vtilde^d) + 2i (\Vtilde^{[a} V^{b]} - V^{[a} \Vtilde^{b]}) \\
& & \pi = {\dright L \over \partial (d \om)} = 4(\Vtilde^{c} V^{c} - V^{c} \Vtilde^{c}) \\
& & \pitilde = {\dright L \over \partial (d \omtilde)} = 4 i (V^{c} V^{c} - \Vtilde^{c} \Vtilde^{c}) \label{pi5}
\ena

The $\star$-Hamiltonian is therefore
\eqa
& & H= dV^a \pi_a + d\Vtilde^a \pitilde_a + d \om^{ab} \pi_{ab} + d\om ~\pi + d \omtilde ~ \pitilde  - L = \nonumber\\
& & ~~~~~~~ dV^a \Phi_a + d\Vtilde^a \Phitilde_a + d \om^{ab} \Phi_{ab} + d\om ~\Phi + d \omtilde ~ \Phitilde + \nonumber\\
& & ~~~~~~~ + [\om^{c[a} \om^{b]}_{~~c} + i (\om^{ab} \om + \om\om^{ab}) + {i \over 2} (\om_{ef} \omtilde + \omtilde \om_{ef}) \epsilon^{abef}] \cdot  \nonumber \\
& & ~~~~~~~ \cdot (V^c V^d - \Vtilde^c \Vtilde^d)\epsilon_{abcd} + 2i (\Vtilde_a V_b-V_a \Vtilde_b)] + \nonumber \\
& & ~~~~~~~ + 4({i \over 8} \om^{ab} \om_{ab} -i \omtilde \omtilde +i \om \om) (\Vtilde^c V_c - V^c \Vtilde_c)- \nonumber \\
& & ~~~~~~~ - 4i ({i \over 16} \om^{ab} \om^{cd} \epsilon_{abcd} - i \om \omtilde - i \omtilde \om)(V^c V_c - \Vtilde^c \Vtilde_c)  \label{starHgravity}
\ena
where the constraints $\Phi$ are defined by subtracting the right-hand sides from the left-hand sides in 
(\ref{pi1}) - (\ref{pi5}).

\subsection{Construction of $\star$-gauge generators}

We apply now the algorithm for the search of symmetry generators. We start from the first class 2--forms $\pi_{ab}$, $\pi$, $\pitilde$. They commute with all the constraints $\Phi$, and are therefore good candidates for the $F$ part of the gauge generator $\Gbb$ in (\ref{generatorG}).  Their Poisson bracket with $H$ is given by:
 \eqa
 & & {\dright H \over \partial  \om^{ab}} =  \pi_{c[a} \om^c_{~~b]} + \om_{[b}^{~~c} \pi_{a] c}+i ( \om \pi_{ab} 
 -\pi_{ab} \om)+ {i\over 2} (\omtilde \pi^{cd} - \pi^{cd} \omtilde ) \epsilon_{abcd} + \nonumber \\
 & & ~~~~~~~~~+ {i\over 8}(\om_{ab} \pi - \pi \om_{ab}) + {i\over 16} (\pitilde \om^{cd} - \om^{cd} \pitilde ) \epsilon_{abcd} \\
   & &  {\dright H \over \partial  \om} = i(\om^{ab} \pi_{ab} - \pi_{ab} \om^{ab} + \om \pi - \pi \om + \omtilde \pitilde -\pitilde \omtilde) \\
   & &  {\dright H \over \partial  \omtilde}=  {i\over 2} (\om_{ab} \pi_{cd} - \pi_{cd} \om_{ab})\epsilon^{abcd} + i (\om \pitilde- \pitilde \om - \omtilde \pi + \pi \omtilde)
\ena
Thus, if we take $d\epsi_{ab} \pi^{ab} + d\epsi~ \pi + d \epsitilde~ \pitilde$ as the $F$ part of the
symmetry generator $\Gbb$,  the second condition in (\ref{conditions}) is satisfied when the $G$ part is the sum of the right-hand sides in the above equations. However we have still to satisfy 
the first condition in (\ref{conditions}). This we achieve by adding to the $G$ part a combination of constraints.
The result is the $\star$--$GL(2,C)$ gauge generator:

\eqa
& & \Gbb = d\epsi_{ab} \pi^{ab} + d\epsi~ \pi + d \epsitilde~ \pitilde + \nonumber \\ & &  ~~~+{1\over 2} \epsi_{ab} (V^a \pi^b+\pi^b V^a) + {i \over 4} \epsi^{cd} (\Vtilde^b \pi^a - \pi^a \Vtilde^b) \epsilon_{abcd} + \nonumber \\
& & ~~~+ {1 \over 2} \epsi_{ab} (\Vtilde^a \pitilde^b+\pitilde^b \Vtilde^a) + {i \over 4} \epsi^{cd} (V^b \pitilde^a - \pitilde^a V^b) \epsilon_{abcd} - \nonumber \\
& & ~~~ - \epsi^{ca} (\om^{b}_{~~c} \pi_{ab}+  \pi_{ab} ~\om^{b}_{~~c} )- i \epsi^{ab} (\om \pi_{ab} - \pi_{ab} \om)
- {i \over 2} \epsi_{cd} (\omtilde  \pi_{ab} - \pi_{ab} \omtilde ) \epsilon^{abcd} -\nonumber \\
& & ~~~ - {i \over 8} \epsi_{ab} (\om^{ab} \pi - \pi \om^{ab}) + {i \over 16} \epsi^{cd} (\om^{ab} \pitilde - \pitilde \om^{ab}) \epsilon_{abcd} - \nonumber \\
& & ~~~ -i \epsi (V^a \pi_a - \pi_a V^a) + \epsitilde ( \Vtilde^a \pi_a + \pi_a \Vtilde^a )- i \epsi ( \Vtilde^a \pitilde_a - \pitilde_a \Vtilde^a ) + \epsitilde (V^a \pitilde_a + \pitilde_a V^a)- \nonumber \\
& & ~~~  - i \epsi(\om^{ab} \pi_{ab} - \pi_{ab} \om^{ab}) - {i \over 2} \epsitilde (\om_{ab} \pi_{cd} -
 \pi_{cd} \om_{ab})\epsilon^{abcd} \nonumber \\
& & ~~~ - i\epsi (\om \pi - \pi \om + \omtilde \pitilde - \pitilde \omtilde ) + i \epsitilde(\omtilde \pi - \pi \omtilde - \om \pitilde + \pitilde \om)
\ena

\noi By simple inspection one can verify that indeed it generates the $\star$-$GL(2,C)$ symmetry variations
listed in (\ref{NCvariations1})-(\ref{NCvariations2}). Moreover, we can now obtain also the 
$\star$-$GL(2,C)$ symmetry variations of the momenta:
\eqa
& & \delta_\epsi \pi_a= {\dright \Gbb \over \partial  V^{a}} = {1 \over 2} (\pi^b \epsi_{ab} + \epsi_{ab} \pi^b) 
-{i \over 4} (\pitilde^b  \epsi^{cd} - \epsi^{cd} \pitilde^b)\epsilon_{abcd} - \nonumber \\
& & ~~~~~~~~~~~~~~~~~~~ - i (\pi_a \epsi -  \epsi \pi_a )+ \pitilde_a \epsitilde + \epsitilde \pitilde_a \\
& & \delta_\epsi \pitilde_a= {\dright \Gbb \over \partial  \Vtilde^{a}} = {1 \over 2} (\pitilde^b \epsi_{ab} + \epsi_{ab} \pitilde^b) 
-{i \over 4} (\pi^b  \epsi^{cd} - \epsi^{cd} \pi^b)\epsilon_{abcd} + \nonumber \\
& & ~~~~~~~~~~~~~~~~~~~ + \pi_a \epsitilde +  \epsitilde \pi_a - i(\pitilde_a \epsi - \epsi \pitilde_a) \\
& & \delta_\epsi \pi_{ab} =  {\dright \Gbb \over \partial  \om^{ab}} =-\pi_{c[a} \epsi_{b]}^{~~c}- \epsi_{c[b} \pi_{a]}^{~~c}
-{i \over 8} (\pi \epsi_{ab} - \epsi_{ab} \pi)+{i \over 16} (\pitilde \epsi^{cd} - \epsi^{cd} \pitilde)\epsilon_{abcd}
\nonumber\\
& &~~~~~~~~~~~~~~~~~~~-i (\pi_{ab} \epsi - \epsi \pi_{ab}) - {i\over 2} (\pi^{cd} \epsitilde - \epsitilde \pi^{cd}) \epsilon_{abcd} \\
& & \delta_\epsi \pi =  {\dright \Gbb \over \partial  \om} = -i(\pi_{ab} \epsi^{ab} - \epsi^{ab} \pi_{ab}) - i(\pi\epsi -\epsi\pi+\pitilde \epsitilde -\epsitilde\pitilde) \\
& & \delta_\epsi \pitilde =  {\dright \Gbb \over \partial  \omtilde} = -{i\over 2} (\pi^{ab} \epsi^{cd} - \epsi^{cd} \pi^{ab})\epsilon_{abcd}  + i(\pi\epsitilde -\epsitilde\pi  - \pitilde \epsi + \epsi \pitilde)
\ena
Finally, using the above variations of the momenta and the variations of the fields given in subsection 4.2.2, it is straightforward to check that the $\Phi$ constraints defined in (\ref{starHgravity}) are weakly preserved.

\sect{Conclusions}

In this paper we have developed a covariant hamiltonian formalism for noncommutative
geometric theories, where noncommutativity is implemented via an abelian twist.
In this canonical framework we have introduced $\star$-Poisson brackets, and the notion
of $\star$-canonical gauge generator, providing also an algorithm for its construction.
This is of relevance when studying the (local or global) symmetries of a geometric theory
like gravity or supergravity, and allows to generalize Noether theorems to a NC setting. 

The NC hamiltonian formalism presented here can be directly applied to supergravity theories, whose covariant symmetry generators have been constructed in \cite{C2020}. The use of noncanonical symplectic structures (see for ex. \cite{NCS}) could provide another interesting generalization
of the covariant hamiltonian setting, and of its NC version.

 \section*{Acknowledgements}
 
We thank Paolo Aschieri and Chiara Pagani for useful discussions. We acknowledge partial support from INFN, CSN4, Iniziativa Specifica GSS.  This research has a financial 
support from Universit\`a del Piemonte Orientale.

\appendix

\sect{Twist differential geometry}

The noncommutative deformation of gravity considered in this paper (and introduced in ref.s  \cite{AC1,AC2})
relies on the existence of an associative $\star$-product between
functions and more generally an associative $\westar$ exterior product between forms, 
satisfying the following properties:
\sk
\noi
$\bullet~~$ \noi Compatibility with the exterior differential:
\eq
d(\tau\wedge_\star \tau')=d(\tau)\wedge_\star \tau'=\tau\wedge_\star
d\tau'
\en
$\bullet~~$ Graded cyclicity of the integral:
        \eq
       \int \tau \westar \tau' =  (-1)^{deg(\tau) deg(\tau')}\int \tau' \westar \tau\label{cycltt'}
       \en
      \noi with $deg(\tau) + deg(\tau')$ = $d$ = dimension of the spacetime
      manifold, and where $\tau$ and $\tau'$ have compact support
      (i.e.  we require (\ref{cycltt'}) to hold up to
      boundary terms).
\sk
\noi $\bullet~~$ Compatibility with complex conjugation:
\eq
       (\tau \westar \tau')^* =   (-1)^{deg(\tau) deg(\tau')} \tau'^* \westar \tau^*
\en
{}Following \cite{AC1}  we describe here a class of twists whose associated $\star$-products
 have all these properties.
This class includes 
the Groenewold-Moyal $\star$-product
\begin{equation}
f\star g = \mu \big{\{} e^{\frac{i}{2}\theta^{\rho\sigma}\partial_\rho \otimes\partial_\sigma}
f\otimes g \big{\}} , \label{MWstar}
\end{equation}
where the map $\mu$  is the usual pointwise
multiplication: $\mu (f \otimes g)= fg$, and $\theta^{\rho\sigma}$ is a constant
antisymmetric matrix.

\sk

\noi{\bf Abelian Twists}
\sk
\noi Let $\Xi$ be the linear space of smooth vector fields on a smooth manifold $M$, and $U\Xi$ its
universal enveloping algebra. A twist  ${\cal F} \in U\Xi \otimes U\Xi$
defines the associative $\star$-product
\begin{eqnarray}
f\star g &=& \mu \big{\{} {\cal F}^{-1} f\otimes g \big{\}}
\end{eqnarray}
\noi  where the map $\mu$  is the usual pointwise
multiplication: $\mu (f \otimes g)= fg$. The product associativity relies on the defining properties of the twist, see for ex. \cite{book}.

   \noi Explicit examples of twist are provided by the so-called abelian twists:
\eq
{\cal F}^{-1}= e^{\frac{i}{2}\theta^{AB}X_A \otimes X_B} \label{Abeliantwist}
\en
where $\{X_A\}$ is a set of mutually commuting vector fields globally
defined on the manifold, and $\theta^{AB}$ is a constant
antisymmetric matrix. The corresponding $\star$-product is in general
position dependent because the vector fields $X_A$ are in general
$x$-dependent. In the special case that there exists a
global coordinate system on the manifold we can consider the
vector fields $X_A={\partial \over \partial x^A}$, corresponding to
the Moyal twist, cf. (\ref{MWstar}):
  \eq
   {\cal F}^{-1}=  e^{\frac{i}{2}\theta^{\rho\sigma}\partial_\rho \otimes\partial_\sigma} \label{Mtwist}
   \en

  \noi {\bf Deformed exterior product}
    \sk

   \noi For abelian twists (\ref{Abeliantwist}), the deformed exterior product between forms is defined as
   \eqa
   & & \tau \westar \tau' \equiv \sum_{n=0}^\infty \left({i \over 2}\right)^n \theta^{A_1B_1} \cdots \theta^{A_nB_n}
   (\ell_{X_{A_1}} \cdots \ell_{X_{A_n}} \tau) \we  (\ell_{X_{B_1}} \cdots \ell_{X_{B_n}} \tau')  \nonumber \\
  & & ~~ = \tau \we \tau' + {i \over 2} \theta^{AB} (\ell_{X_A} \tau) \we (\ell_{X_B} \tau') + {1 \over 2!}  {\left( i \over 2 \right)^2} \theta^{A_1B_1} \theta^{A_2B_2}  (\ell_{X_{A_1}} \ell_{X_{A_2}} \tau) \we
 (\ell_{X_{B_1}} \ell_{X_{B_2}} \tau') + \cdots \nonumber 
  \label{defwestar}
  \ena
       \noi where the commuting tangent vectors $X_A$ act on forms via the Lie derivatives
       ${\ell}_{X_A} $. 
     This product is associative, and the above formula holds also for $\tau$ or $\tau'$ being a $0$-form (i.e. a function). 
 
\sk
\noi {\bf Exterior derivative}
        \sk
         \noi The exterior derivative satisfies the usual (graded) Leibniz rule,
         since it commutes with the Lie derivative:
        \eqa
        & & d (f \star g) = df \star g + f \star dg \\
        & & d(\tau \westar \tau') = d\tau \westar \tau'  + (-1)^{deg(\tau)} ~\tau \westar d\tau'
        \ena

\sk

       \noi {\bf Integration: graded cyclicity} \nopagebreak
        \sk
        \noi If we consider an abelian twist (\ref{Abeliantwist})
        given by globally defined commuting vector fields $X_A$,
        then the usual integral is cyclic under the $\star$-exterior
        products of forms, i.e. equation (\ref{cycltt'}) holds up to boundary terms.
       Indeed:
\eq       \int \tau \westar \tau' =    \int \tau \wedge \tau'=
(-1)^{deg(\tau) deg(\tau')}\int \tau' \wedge \tau=
(-1)^{deg(\tau) deg(\tau')}\int \tau' \westar \tau
\en
up to boundary terms. For example at first order in $\theta$,
\eq
\int \tau \westar \tau' =    \int \tau \wedge \tau'-{i\over
 2}\theta^{AB}\int{\cal L}_{X_A}(\tau\wedge {\cal L}_{X_B}\tau')
=
\int \tau \wedge \tau'-{i\over
 2}\theta^{AB}\int d {i}_{X_A}(\tau\wedge {\cal L}_{X_B}\tau')
\en
where we used the Cartan formula ${\cal L}_{X_A}=di_{X_A}+i_{X_A}d$.
\sk       \noi {\bf Complex conjugation}
    \sk
        \noi If we choose real fields $X_A$ in the definition of the
        twist (\ref{Abeliantwist}),  it is immediate to verify that:
        \eq
        (f \star g)^* = g^* \star f^*\label{starfg*}
        \en
        \eq
        (\tau \westar \tau')^* =   (-1)^{deg(\tau) deg(\tau')} \tau'^* \westar \tau^*\label{startt*}
        \en
        since sending $i$ into $-i$ in the twist (\ref{Mtwist}) amounts to send $\theta^{AB}$ into
        $-\theta^{AB} = \theta^{BA}$, i.e. to exchange the
        order of the factors in the $\star$-product.

\sect{Gamma matrices in $d=4$}

We summarize in this Appendix our gamma matrix conventions in $d=4$.

\eqa
& & \eta_{ab} =(1,-1,-1,-1),~~~\{\ga_a,\ga_b\}=2 \eta_{ab},~~~[\ga_a,\ga_b]=2 \ga_{ab}, \\
& & \ga_5 \equiv i \ga_0\ga_1\ga_2\ga_3,~~~\ga_5 \ga_5 = 1,~~~\epsi_{0123} = - \epsi^{0123}=1, \\
& & \ga_a^\dagger = \ga_0 \ga_a \ga_0, ~~~\ga_5^\dagger = \ga_5 \\
& & \ga_a^T = - C \ga_a C^{-1},~~~\ga_5^T = C \ga_5 C^{-1}, ~~~C^2 =-1,~~~C^\dagger=C^T =-C
\ena

\subsection{Useful identities}

\eqa
 & &\ga_a\ga_b= \ga_{ab}+\eta_{ab}\\
 & & \ga_{ab} \ga_5 = {i \over 2} \epsilon_{abcd} \ga^{cd}\\
 & &\ga_{ab} \ga_c=\eta_{bc} \ga_a - \eta_{ac} \ga_b -i \epsi_{abcd}\ga_5 \ga^d\\
 & &\ga_c \ga_{ab} = \eta_{ac} \ga_b - \eta_{bc} \ga_a -i \epsi_{abcd}\ga_5 \ga^d\\
 & &\ga_a\ga_b\ga_c= \eta_{ab}\ga_c + \eta_{bc} \ga_a - \eta_{ac} \ga_b -i \epsi_{abcd}\ga_5 \ga^d\\
 & &\ga^{ab} \ga_{cd} = -i \epsi^{ab}_{~~cd}\ga_5 - 4 \de^{[a}_{[c} \ga^{b]}_{~~d]} - 2 \de^{ab}_{cd}\\
& & Tr(\ga_a \ga^{bc} \ga_d)= 8~ \de^{bc}_{ad} \\
& & Tr(\ga_5 \ga_a \ga_{bc} \ga_d) = -4 ~\epsi_{abcd} \\
& & Tr(\ga^{rs} \ga_a \ga_{bc} \ga_d)=4(-2 \de^{rs}_{cd} \eta_{ab} + 2 \de^{rs}_{bd} \eta_{ac} - 3! \de^{rse}_{abc} \eta_{ed}) \\
& & Tr(\ga_5 \ga^{rs} \ga_a \ga_{bc} \ga_d)=
4(-i \eta_{ab} \epsi^{rs}_{~~cd} + i \eta_{ac} \epsi^{rs}_{~~bd} + 2i \epsi_{abc}^{~~~e} \de^{rs}_{ed})
 \ena
\sk
\noi where
$\delta^{ab}_{cd} \equiv \frac{1}{2}(\delta^a_c\delta^b_d-\delta^b_c\delta^a_d)$, $\delta^{rse}_{abc} \equiv  {1 \over 3!} (\de^r_a \de^s_b \de^e_c$ + 5 terms), 
and indices antisymmetrization in square brackets has total weight $1$.

\vfill\eject
\end{document}